\documentclass[acmtocl]{acmtrans2m}

\newtheorem{theorem}{Theorem}[section]

\newdef{definition}[theorem]{Definition}
\newdef{remark}[theorem]{Remark}

\title{Analysis of the postulates produced by Karp's Theorem.}
\author{JERRALD MEEK}

\begin{abstract}
This is the final article in a series of four articles.  Richard Karp has proven that a deterministic polynomial time solution to K-SAT will result in a deterministic polynomial time solution to all \textit{NP-Complete} problems.  However, it is demonstrated that a deterministic polynomial time solution to any \textit{NP-Complete} problem does not necessarily produce a deterministic polynomial time solution to all \textit{NP-Complete} problems.
\end{abstract}


\begin{document}

\begin{bottomstuff}
People who wish to remain anonymous have offered comments and suggestions which have
improved this work. The author wishes to express his appreciation for their assistance.

	\begin{flushright}
		Jerrald Meek Copyright \copyright 2008
	\end{flushright}
\end{bottomstuff}

\maketitle

\section{Introduction.}
The present author has previously shown that a \textit{NP-complete} problem is solvable in deterministic polynomial time by a search algorithm if and only if a polynomial search partition can be found in polynomial time.\par

Additionally, it has been shown that some instances of the \textit{0-1-Knapsack} problem do not have a deterministic polynomial time solution, unless the SAT problem has a deterministic polynomial time solution.\par

This is the final article in a series of four, wherein the purpose will be to finally determine that the SAT problem has no deterministic polynomial time solution.  It will then become clear that the \textit{0-1-Knapsack} problem, and any similar \textit{NP-complete} problem which depends on SAT for a deterministic polynomial time solution, cannot be solved in deterministic polynomial time.

\section{Preliminaries.}
Previously the author has proven the following theorems, which will be assumed true in this article.\par

\newtheorem{thm_opt}{Theorem 4.4 from \textit{P is a proper subset of NP}. \protect\cite{meek1}}[section]
\begin{thm_opt}
\textbf{\emph{P = NP Optimization Theorem.}}\par

The only deterministic optimization of a \textit{NP-complete} problem that could prove \textit{P = NP} would be one that can always solve a \textit{NP-complete} problem by examining no more than a polynomial number of input sets for that problem.
\end{thm_opt}

\newtheorem{thm_part}[thm_opt]{Theorem 5.1 from \textit{P is a proper subset of NP}. \protect\cite{meek1}}
\begin{thm_part}
\textbf{\emph{P = NP Partition Theorem.}}\par

The only deterministic search optimization of a \textit{NP-complete} problem that could prove \textit{P = NP} would be one that can always find a representative polynomial search partition by examining no more than a polynomial number of input sets from the set of all possible input sets.
\end{thm_part}

\newtheorem{thm_rnd}[thm_opt]{Theorem 5.1 from \protect\cite{meek2}}
\begin{thm_rnd}
\begin{tabbing}
1234567890\=\+\kill
	\\
	\textit{Analysis of the deterministic polynomial time solvability}\\
	\textit{of the 0-1-Knapsack problem}.
\end{tabbing}
\noindent \textbf{\emph{Knapsack Random Set Theorem.}}\par

Deterministic Turing Machines cannot exploit a random relation between the elements of $S$ to produce a polynomial time solution to the \textit{Knapsack} problem.
\end{thm_rnd}

\newtheorem{thm_comp}[thm_opt]{Theorem 6.1 from \protect\cite{meek2}}
\begin{thm_comp}
\begin{tabbing}
1234567890\=\+\kill
	\\
	\textit{Analysis of the deterministic polynomial time solvability}\\
	\textit{of the 0-1-Knapsack problem}.
\end{tabbing}
\noindent \textbf{\emph{Knapsack Composition Theorem.}}\par

Compositions of $M$ cannot be relied upon to always produce a deterministic polynomial time solution to the \textit{0-1-Knapsack} problem.
\end{thm_comp}

\newtheorem{thm_m}[thm_opt]{Theorem 6.2 from \protect\cite{meek2}}
\begin{thm_m}
\begin{tabbing}
1234567890\=\+\kill
	\\
	\textit{Analysis of the deterministic polynomial time solvability}\\
	\textit{of the 0-1-Knapsack problem}.
\end{tabbing}
\noindent \textbf{\emph{Knapsack $M$ Quality Reduction Theorem.}}\par

Any quality of $M$ that could be used to find a composition of $M$ within $S$ would be equivalent to finding all compositions of $M$.
\end{thm_m}

\newtheorem{thm_set}[thm_opt]{Theorem 7.1 from \protect\cite{meek2}}
\begin{thm_set}
\begin{tabbing}
1234567890\=\+\kill
	\\
	\textit{Analysis of the deterministic polynomial time solvability}\\
	\textit{of the 0-1-Knapsack problem}.
\end{tabbing}
\noindent \textbf{\emph{Knapsack Set Quality Theorem.}}\par

Using any quality of the elements of $S$ to solve the \textit{0-1-Knapsack} problem will be no less complex than the standard means of solving the \textit{0-1-Knapsack} problem.
\end{thm_set}

The definition of the \textit{0-1-Knapsack} problem used in this article will be based off of that used by Horowitz and Sahni \protect\cite{horowitz}.
\noindent \begin{enumerate}
	\item Let $S$ be a set of real numbers with no two identical elements.
	\item Let $r$ be the number of elements in $S$.
	\item Let $\delta$ be a set with $r$ elements such that
		\[ {\delta}_i \in \left\{ {0, 1} \right\} \leftarrow 1 \leq i \leq r \]
	\item Let $M$ be a real number.
\end{enumerate}

Then
\[ \sum_{i=1}^r {\delta}_i S_i = M \]

Find a variation of $\delta$ that causes the expression to evaluate \textit{true}.

\section{Assumptions produced by Karp's Theorem.}

\newtheorem{thm_karp}{Theorem 3 from \textit{Reducibility among combinatorial problems}. \protect\cite[p. 93]{karp}}[section]
\begin{thm_karp}

\noindent The language $L$ is $(\underline{polynomial})$ $\underline{complete}$ if

\begin{tabbing}
	12345678\=\kill
	\>a) $L\in NP$ \\
	and\>b) SATISFIABILITY $\propto L$ \\
\end{tabbing}

\noindent Either all complete languages are in $P$, or none of them are. The former alternative holds if and only if \textit{P = NP}.
\end{thm_karp}

Karp's Theorem is often interpreted to mean that if any \textit{NP-Complete} problem is discovered to be solvable in deterministic polynomial time, then all \textit{NP-Complete} problems will be solvable in deterministic polynomial time.  This common interpretation is based off of two postulates:
\begin{enumerate}
	\item K-SAT can be reduced to any \textit{NP-Complete} problem, and therefore any \textit{NP-Complete} problem can be solved in deterministic polynomial time if the underlying K-SAT problem has a fast solution.
	\item A deterministic polynomial time solution to some \textit{NP-Complete} problem will ultimately provide a deterministic polynomial time solution to the underlying K-SAT problem.
\end{enumerate}

In this article the author has no argument with the first postulate regarding Karp's Theorem.  Karp has satisfactorily demonstrated that K-SAT can be reduced to any \textit{NP-Complete} problem.  It was Richard Karp who suggested that a deterministic polynomial time solution to K-SAT would prove $P$ = \textit{NP}; the present author agrees.\par

The true purpose of this article is to challenge the second postulate.  The idea that a deterministic polynomial time solution to any \textit{NP-Complete} problem will magically produce a proof that $P$ = \textit{NP} is often assumed to be implied by Karp's Theorem; however it is not so implied, and this idea is completely misguided.

\section{A \textit{NP-Complete} problem with a deterministic polynomial time solution.}

It has been previously demonstrated by the present author that the process of converting a decimal number into a binary number can be represented as a form of the \textit{0-1-Knapsack} problem, and therefore is \textit{NP-Complete}.  However, this particular \textit{NP-Complete} problem does have a deterministic polynomial time solution.\par

\subsection{Formal proof that base conversion is a \textit{NP-Complete} problem.}

Shortly after the publication of the first version of this article, an arXiv reader who has been kind enough to allow the quoting of his e-mail sent the author the following message:\par

\begin{quotation}
I saw your ArXiv article 0808.3222.  There's a fundamental misunderstanding in that paper.  Just because base conversion is a *special case* of an NP-complete problem doesn't mean that base conversion is NP-complete.  You have the relation backwards.  If you could show that knapsack was a special case of base conversion, *then* base conversion would be NP-complete.

\begin{flushright}
Timothy Chow; 26 August, 2008
\end{flushright}
\end{quotation}

\vspace{2ex}
The author realizes that in article 2 \protect\cite{meek2} of this series, the author did not necessarily give a complete formal proof that base conversion is \textit{NP-Complete}.  Furthermore, this objection is actually a reasonable argument, which the author should have foreseen.  Therefore, a proof will be demonstrated here, wherein the method of Richard Karp will be followed.  First the \textit{0-1-Knapsack} problem will be proven \textit{NP-Complete} by reduction of K-SAT, to the Knapsack problem.

Karp actually proved the Knapsack problem to be \textit{NP-Complete} by the following progression:
\begin{enumerate}
	\item SATISFIABILITY $\propto$ SATISFIABILITY WITH AT MOST 3 LITERALS PER CLAUSE
	\item SATISFIABILITY WITH AT MOST 3 LITERALS PER CLAUSE $\propto$ CHROMATIC NUMBER
	\item CHROMATIC NUMBER $\propto$ EXACT COVER
	\item EXACT COVER $\propto$ KNAPSACK
\end{enumerate}

\vspace{1ex}
In the following proof Karp's rules are adhered to, however for the sake of brevity we will reduce SAT directly to 0-1-Knapsack.

\vspace{1ex}
\begin{proof}
Assume:
\begin{itemize}
	\item $S = \left\langle {x, y, z | x \in R, y \in R, z \in R} \right\rangle$
	\item $M \in R$
	\item $a = \left\langle
		\begin{array}{l}
			{ \left[ {0x + 0y + 0z = M} \right], \left[ {x + 0y + 0z = M} \right], \left[ {0x + y + 0z = M} \right], \left[ {0x + 0y + z = M} \right], } \\
			{ \left[ {x + y + 0z = M} \right], \left[ {x + 0y + z = M} \right], \left[ {0x + y + z = M} \right], \left[ {x + y + z = M} \right] } \\
		\end{array} \right\rangle $
\end{itemize}

\vspace{1ex}
\noindent An 8-SAT problem is
\[ a_1 \vee a_2 \vee a_3 \vee a_4 \vee a_5 \vee a_6 \vee a_7 \vee a_8 \]

\noindent Substitute the elements of $a$ for their assigned values
\[ \left[ {0x + 0y + 0z = M} \right] \vee \left[ {x + 0y + 0z = M} \right] \vee \left[ {0x + y + 0z = M} \right] \vee \]
\[ \left[ {0x + 0y + z = M} \right] \vee \left[ {x + y + 0z = M} \right] \vee \left[ {x + 0y + z = M} \right] \vee \]
\[ \left[ {0x + y + z = M} \right] \vee \left[ {x + y + z = M} \right] \]

\noindent Convert the literals into summations
\[ \left[ {\sum_{i = 1}^3 \left\{ {0, 0, 0} \right\}_i S_i = M} \right] \vee \left[ {\sum_{i = 1}^3 \left\{ {1, 0, 0} \right\}_i S_i = M} \right] \vee \left[ {\sum_{i = 1}^3 \left\{ {0, 1, 0} \right\}_i S_i = M} \right] \vee \]
\[ \left[ {\sum_{i = 1}^3 \left\{ {0, 0, 1} \right\}_i S_i = M} \right] \vee \left[ {\sum_{i = 1}^3 \left\{ {1, 1, 0} \right\}_i S_i = M} \right] \vee \left[ {\sum_{i = 1}^3 \left\{ {1, 0, 1} \right\}_i S_i = M} \right] \vee \]
\[ \left[ {\sum_{i = 1}^3 \left\{ {0, 1, 1} \right\}_i S_i = M} \right] \vee \left[ {\sum_{i = 1}^3 \left\{ {1, 1, 1} \right\}_i S_i = M} \right] \]

The above logical disjunction can more easily be restated similarly to the Horowitz and Sahni definition of the \textit{0-1-Knapsack} problem
\[ \sum_{i = 1}^3 \delta_i S_i = M \]

\noindent Does a variation of $\delta$ exist which makes this expression \textit{true}?
\end{proof}

Now that we have a formal proof that the \textit{0-1-Knapsack} problem is a member of the \textit{NP-Complete} class.  The same method can be used for base conversion.

\vspace{1ex}
\begin{proof}
Assume:
\begin{itemize}
	\item $S = \left\langle {1, 2, 4} \right\rangle $
	\item $M = 6$
	\item $a = \left\langle
		\begin{array}{l}
			{ \left[ {0(1) + 0(2) + 0(4) = 6} \right], \left[ {1 + 0(2) + 0(4) = 6} \right], \left[ {0(1) + 2 + 0(4) = 6} \right], \left[ {0(1) + 0(2) + 4 = 6} \right], } \\
			{ \left[ {1 + 2 + 0(4) = 6} \right], \left[ {1 + 0(2) + 4 = 6} \right], \left[ {0(1) + 2 + 4 = 6} \right], \left[ {1 + 2 + 4 = 6} \right] }
		\end{array} \right\rangle $
\end{itemize}

\vspace{1ex}
\noindent An 8-SAT problem is
\[ a_1 \vee a_2 \vee a_3 \vee a_4 \vee a_5 \vee a_6 \vee a_7 \vee a_8 \]

\noindent Substitute the elements of $a$ for their assigned values
\[ \left[ {0(1) + 0(2) + 0(4) = 6} \right] \vee \left[ {1 + 0(2) + 0(4) = 6} \right] \vee \left[ {0(1) + 2 + 0(4) = 6} \right] \vee \]
\[ \left[ {0(1) + 0(2) + 4 = 6} \right] \vee \left[ {1 + 2 + 0(4) = 6} \right] \vee \left[ {1 + 0(2) + 4 = 6} \right] \vee \]
\[ \left[ {0(1) + 2 + 4 = 6} \right] \vee \left[ {1 + 2 + 4 = 6} \right] \]

\noindent Convert the literals into summations
\[ \left[ {\sum_{i = 1}^3 \left\{ {0, 0, 0} \right\}_i S_i = 6} \right] \vee \left[ {\sum_{i = 1}^3 \left\{ {1, 0, 0} \right\}_i S_i = 6} \right] \vee \left[ {\sum_{i = 1}^3 \left\{ {0, 1, 0} \right\}_i S_i = 6} \right] \vee \]
\[ \left[ {\sum_{i = 1}^3 \left\{ {0, 0, 1} \right\}_i S_i = 6} \right] \vee \left[ {\sum_{i = 1}^3 \left\{ {1, 1, 0} \right\}_i S_i = 6} \right] \vee \left[ {\sum_{i = 1}^3 \left\{ {1, 0, 1} \right\}_i S_i = 6} \right] \vee \]
\[ \left[ {\sum_{i = 1}^3 \left\{ {0, 1, 1} \right\}_i S_i = 6} \right] \vee \left[ {\sum_{i = 1}^3 \left\{ {1, 1, 1} \right\}_i S_i = 6} \right] \]

The above logical disjunction can more easily be restated similarly to the Horowitz and Sahni definition of the \textit{0-1-Knapsack} problem
\[ \sum_{i = 1}^3 \delta_i S_i = 6 \]

\noindent Does a variation of $\delta$ exist which makes this expression \textit{true}?
\end{proof}

The point here is that K-SAT is reducible to the \textit{0-1-Knapsack} problem regardless of what numbers are placed into set $S$ and value $M$.  It is here demonstrated that finding the base 2 digits of 6 is a \textit{NP-Complete} problem.\par

The only variation of $\delta$, which evaluates \textit{true}, is the ordered set $\delta = \left\langle {0, 1, 1} \right\rangle $.  The base 2 representation of 6 is 110.  Notice that the left most digit in 110 represents a value of 4, while the center digit represents a value of 2.  Therefore, 110 represents 4 + 2 = 6.  Likewise, $\delta = \left\langle {0, 1, 1} \right\rangle $ when $S = \left\langle {1, 2, 4} \right\rangle $ represents 2 + 4 = 6.
\[ 4 + 2 = 2 + 4 \]

We already know that this particular \textit{Knapsack} problem is solvable in deterministic polynomial time, as demonstrated in article 2 \protect\cite{meek2}, but it is also solvable in non-deterministic polynomial time and is then still a member of \textit{NP}.  Therefore, our ``special case'' of the \textit{knapsack} problem meets both of Karp's requirements to be an element of the \textit{NP-Complete} class.

\subsection{How some \textit{NP-Complete} problems have fast solutions.}

The reason that the problem of converting a decimal number to a binary number can be solved in deterministic polynomial time is because there is a special relationship between the elements of $S$ in this particular \textit{0-1-Knapsack} problem.  This relationship can therefore be exploited to determine which literal of the underlying K-SAT problem has a \textit{true} value.  See \protect\cite[sec 5.1]{meek2}.\par

\vspace{1ex}
\begin{proof}
Assume the following:
\begin{itemize}
	\item The set S has r elements.
	\item There are $r^2$ subsets of $S$.
	\item Some special relationship between the elements of $S$ allows for a fast determination of one subset of $S$ which sums to $M$.  That same relationship also allows a fast determination in the event that no subset of $S$ sums to $M$.
	\item Let $\delta$ be a set with $r$ elements such that
		\[ {\delta}_i \in \left\{ {0, 1} \right\} \leftarrow 1 \leq i \leq r \]
	\item Let $M$ be a real number.
\end{itemize}

\vspace{1ex}
\noindent Then
\[ \sum_{i=1}^r {\delta}_i S_i = M \]

\noindent can be represented as

\[ \left[0 = M\right] \vee \left[S_1 = M\right] \vee \left[S_2 = M\right] \vee \left[S_3 = M\right] \vee \ldots \left[S_r = M\right] \vee \]
\[ \left[S_1 + S_2 = M\right] \vee \left[S_1 + S_3 = M\right] \vee \ldots \]

The above K-SAT problem is extended until there is exactly one literal for each subset of $S$.  To determine the truth-value of the entire expression it is only necessary to determine that any one literal is \textit{true}, or that all literals are \textit{false}.\par

Because of the special relationship between the elements of $S$, it is possible to quickly find one literal that is \textit{true} if one exists \protect\cite{meek2}.  It is also the case that if an attempt to find such a literal fails, then it can be quickly determined that all laterals are \textit{false}.\par

The fast solution to this problem is produced by the fact that it is not necessary to determine the truth-value of each and every literal individually.  If it were the case that each literal had to be individually determined, then the \textit{P = NP Optimization Theorem} would not allow the optimization to process in deterministic polynomial time (see the discussion of this in section 5 below).\par

It is then easy to see that the deterministic polynomial time solution to this problem cannot produce a fast solution to the underlying K-SAT problem.  Infact, if the optimization did produce a solution to the underlying K-SAT problem, then the solution could not run in deterministic polynomial time.\par

Instead, this optimized solution simply produces a fast identification of one \textit{true} literal in the K-SAT problem.  This ability to quickly identify a \textit{true} literal in the K-SAT problem cannot be transferred to other \textit{NP-Complete} problems, because the fast solution is dependant upon a special condition of a particular instance of the \textit{0-1-Knapsack} problem, which will not exist in all other \textit{NP-Complete} problems.
\end{proof}

\subsection{Formalized argument proving no deterministic polynomial time solution from the form of a \textit{NP-Complete} problem for K-SAT where $k \geq 3$.}

\vspace{1ex}
\noindent Assume:
\begin{itemize}
	\item The K-SAT problem associated to the hereinabove described \textit{0-1-Knapsack} problem has $k$ literals where $k = 2^{\left|S\right|} \geq 3$.
	\item $x$ is a set of literals with $k$ elements; each literal has an unknown truth value.
\end{itemize}

\noindent The K-SAT problem in question is
\[ x_1 \vee x_2 \vee x_3 \vee \ldots \vee x_k \]

\vspace{1ex}
\begin{proof}
Assume:
\begin{itemize}
	\item $Q$ = some predictable relationship between the elements of $S$.
	\item $S = \left\{ {x | Q \mapsto x} \right\} $
	\item $M \in R$
	\item $x_2 = T$
	\item $A\left( {y, z} \right)$ is an algorithm which identifies one element of $x$ which has a \textit{true} value by comparing the elements of $y$ with relation $Q$ to the value of $z$.
\end{itemize}

\vspace{1ex}
\noindent Then
\[ A\left( {S, M} \right) \equiv x_2 \]

The optimized \textit{0-1-Knapsack} algorithm uses the elements of $S$ with quality $Q$, and the value of $M$ to identify that one literal has a \textit{true} value.
\end{proof}

\vspace{1ex}
\begin{proof}
Assume:
\begin{itemize}
	\item $Q$ = some predictable relationship between the elements of $S$.
	\item $S = \left\{ {x| x \in R} \right\}$
	\item $M \in R$
	\item $A\left( {y, z} \right)$ is an algorithm which identifies one element of $x$ which has a \textit{true} value by comparing the elements of $y$ with relation $Q$ to the value of $z$.
	\item $\otimes$ represents an undefined result.
\end{itemize}

\vspace{1ex}
\noindent Then
\[ A\left( {S, M} \right) \equiv \otimes \]

The optimized \textit{0-1-Knapsack} algorithm uses the elements of $S$ with quality $Q$, and the value of $M$ to identify that one literal has a \textit{true} value.  However, the quality $Q$ is not associated with the elements of $S$, so the result is undefined.
\end{proof}

\vspace{1ex}
\begin{proof}
Assume:
\begin{itemize}
	\item $Q$ = some predictable relationship between the elements of $S$.
	\item We are given a K-COL problem (graph coloring problem where $k \geq 3$).
	\item $S = \oslash$
	\item $M = \oslash$
	\item $A\left( {y, z} \right)$ is an algorithm which identifies one element of $x$ which has a \textit{true} value by comparing the elements of $y$ with relation $Q$ to the value of $z$.
	\item $\otimes$ represents an undefined result.
\end{itemize}

\vspace{1ex}
\noindent Then
\[ A\left( {S, M} \right) \equiv \otimes \]

The optimized \textit{0-1-Knapsack} algorithm uses the elements of $S$ with relation $Q$, and the value of $M$ to identify that one literal has a \textit{true} value.  However, the set $S$ and value $M$ are not provided in the definition of the K-COL problem, so the result is undefined.
\end{proof}

\begin{theorem}
\textbf{\emph{K-SAT Input Relation Theorem.}}\par

A solution that solves a \textit{NP-Complete} problem in deterministic polynomial time, and relies upon some relationship between the inputs of the problem, does not produce a deterministic polynomial time solution for all instances of K-SAT.
\end{theorem}

\section{Magical solutions not allowed.}
When solving any \textit{NP-Complete} problem we are given the following:
\begin{enumerate}
	\item a formula representing the problem;
	\item the input for the problem;
	\item the set of all possible inputs for the underlying K-SAT.
\end{enumerate}

If some magical algorithm exists that allows someone to solve all \textit{NP-Complete} problems in deterministic polynomial time, then the algorithm is stuck working with one or more of the preceding options.  So far we have ruled out option 3 in article 1 \protect\cite{meek1}, and option 2 is eliminated in article 2 \protect\cite{meek2} and the present work.\par

So we are left with option 1, the formula.  In the case of \textit{NP-Complete} problems, that formula is K-SAT, where $k \geq 3$.  K-SAT can be reduced to any \textit{NP-Complete} problem, and any \textit{NP-Complete} problem can be represented as a form of K-SAT.\par

It will be assumed here that 3-SAT cannot be reduced to 2-SAT.  For further discussion on the complexity of a K-SAT formula, see article 1 \protect\cite{meek1}.\par

\subsection{Fast K-SAT and the rules of arithmetic.}
Shortly after the publication of the first version of this article on arXiv, the author received an e-mail from Lance Fortnow containing a warning about the dificulty of resolving $P$ vs. \textit{NP}.  With the permission of Dr. Fortnow, his statement is reproduced, ``be sure that [this research report will] rule out all possible algorithms making absolutely no assumption about their behavior.''\par

The question is how do we know for sure, even after determining that none of the 3 tools given to us to work with are sufficient, that there isn't some strange number theoretic trick that bypasses all of these limitations?\par

\vspace{1ex}
\begin{proof}
Assume:
\begin{itemize}
	\item $a$, $b$, and $c$ are all literals with unknown truth values.
	\item There exists a base 3-SAT problem $a \vee b \vee c$
	\item $x = \left\langle { \ \ a, \ \ a, \neg a, \ a, \neg a, \neg a, \neg a} \right\rangle $
	\item $y = \left\langle { \ \ b, \neg b, \ \ b, \neg b, \ \ b, \neg b, \neg b} \right\rangle $
	\item $z = \left\langle {\neg c, \ \ c, \ \ c, \neg c. \neg c, \ \ c, \neg c} \right\rangle $
\end{itemize}

We can now create the following 7 additional formulas, which are contingent upon the base formula.
\begin{enumerate}
	\item $x_1 \vee y_1 \vee z_1 \ \ \left[ { \ = \ \ a \vee \ \ b \vee \neg c \ } \right]$
	\item $x_2 \vee y_2 \vee z_2 \ \ \left[ { \ = \ \ a \vee \neg b \vee \ \ c \ } \right]$
	\item $x_3 \vee y_3 \vee z_3 \ \ \left[ { \ = \neg a \vee \ \ b \vee \ \ c \ } \right]$
	\item $x_4 \vee y_4 \vee z_4 \ \ \left[ { \ = \ \ a \vee \neg b \vee \neg c \ } \right]$
	\item $x_5 \vee y_5 \vee z_5 \ \ \left[ { \ = \neg a \vee \ \ b \vee \neg c \ } \right]$
	\item $x_6 \vee y_6 \vee z_6 \ \ \left[ { \ = \neg a \vee \neg b \vee \ \ c \ } \right]$
	\item $x_7 \vee y_7 \vee z_7 \ \ \left[ { \ = \neg a \vee \neg b \vee \neg c \ } \right]$
\end{enumerate}

Notice that the truth-value of our 7 formulas is contingent upon not simply the truth-value of the base 3-SAT formula, but the truth-value of all three literals of the base formula.\par

\begin{table}[h]
\begin{tabular}{|c|c|c|c|c|c|c|c|c|c|c|c|}
	\hline
	 & \multicolumn{3}{c|}{\textbf{Literal}} & \multicolumn{8}{c|}{\textbf{Formula}} \\
	\hline
	 & a & b & c & BASE & 1 & 2 & 3 & 4 & 5 & 6 & 7 \\
	\hline
	1 & T & T & T & T & T & T & T & T & T & T & F \\
	\hline
	2 & T & T & F & T & T & T & T & T & T & F & T \\
	\hline
	3 & T & F & T & T & T & T & T & T & F & T & T \\
	\hline
	4 & F & T & T & T & T & T & T & F & T & T & T \\
	\hline
	5 & T & F & F & T & T & T & F & T & T & T & T \\
	\hline
	6 & F & T & F & T & T & F & T & T & T & T & T \\
	\hline
	7 & F & F & T & T & F & T & T & T & T & T & T \\
	\hline
	8 & F & F & F & F & T & T & T & T & T & T & T \\
	\hline
\end{tabular}
\end{table}

If an algorithm determines that $a$ is \textit{true} without determining the value of $b$ or $c$, then the tautology hood of formulas 3, 5, 6, and 7 are still unknown.\par

As can be seen from the above table, an algorithm that identifies only one literal that is \textit{true}, or alternatively determines that all literals are \textit{false} will not suffice.  Such an algorithm would solve the base problem, but the solution would not transfer to all of the other problems.  We then cannot say that the algorithm would provide a solution for all \textit{NP-Complete} problems.
\end{proof}

So is it possible that some magical algorithm could find the value of all three literals in deterministic polynomial time without performing a search or a partitioned search?

\vspace{1ex}
\begin{proof}
Assume:
\begin{itemize}
	\item The values of $a$, $b$, and $c$ are independent of each other; as is suggested by the above table.
	\item $a \Rightarrow d = 1, \neg a \Rightarrow d = 0$.
	\item $b \Rightarrow e = 2, \neg b \Rightarrow e = 0$.
	\item $c \Rightarrow f = 4, \neg c \Rightarrow f = 0$.
	\item $g \in \left[ {0, 7} \right]$
\end{itemize}

\vspace{1ex}
\noindent Given
\[ d + e + f = g \]

The value of $g$ can be determined from the three variables of unknown value in deterministic polynomial time if and only if the value of $a$, $b$, and $c$ can be determined by some algorithm in deterministic polynomial time.\par

It is easy to determine that in the above example, any Turing Machine, non-deterministic or deterministic, would be confused between the possibilities of eight different equally valid results.
\end{proof}

\subsection{Fast K-SAT and unknown formulas.}
To complicate matters more.  The exact formula is not known in many \textit{NP-Complete} problems.

\begin{proof}
Assume:
\begin{itemize}
	\item $S = \left\langle {1, 2, 3} \right\rangle $
	\item $M = 5$
	\item $x = \left\langle {
		\begin{array}{l}
			\left[ {0 = 5} \right], \left[ {1 = 5} \right], \left[ {2 = 5} \right], \left[ {3 = 5} \right], \\
			\left[ {1 + 2 = 5} \right], \left[ {1 + 3 = 5} \right], \left[ {2 + 3 = 5} \right], \left[ {1 + 2 + 3 = 5} \right] \\
		\end{array} } \right\rangle $
\end{itemize}

\noindent The underlying 8-SAT problem for the above-described \textit{0-1-Knapsack} problem is
\[ x_1 \vee x_2 \vee x_3 \vee x_4 \vee x_5 \vee x_6 \vee x_7 \vee x_8 \]

Notice in the above example, $x_7$ is by necessity a literal with a value of \textit{true} because it is associated with the equation $2 + 3 = 5$, which is a \textit{true} statement under the rules of arithmetic.  Likewise, all other literals in this problem are by necessity literals with a \textit{false} value because they are associated with equations, which are \textit{false} statements under the rules of arithmetic.\par

The \textit{P = NP Optimization Theorem} will not allow a deterministic polynomial time algorithm to examine the equations associated with each literal of a \textit{Knapsack} problem to determine which ones are by necessity \textit{true}, and which are by necessity \textit{false}.  Likewise, other \textit{NP-Complete} problems have literals associated with statements, and each literal is by necessity either \textit{true} or \textit{false}.  It is then the case that for each \textit{NP-Complete} problem there is one and only one input set for the underlying K-SAT problem that is valid.\par

It has already been demonstrated that an algorithm cannot select one out of the $2^k$ possible input sets as the correct one for a K-SAT problem.  It has also been demonstrated that even if an algorithm did select the correct one, then there are $2^k - 1$ other \textit{NP-Complete} problems for which that solution is not correct.\par

We could expect that an algorithm could examine a polynomial number of literals to eliminate some possibilities, but there will still be more than one possible solution.
\end{proof}

\section{Conclusion:  $P \neq$ \textit{NP}}

The \textit{P = NP Optimization Theorem} eliminates a search algorithm by exhaustion as a polynomial time solution to a \textit{NP-Complete} problem.\par

The \textit{P = NP Partition Theorem} eliminates a search algorithm to produce a polynomial size search partition which can be used in a polynomial time solution to a \textit{NP-Complete} problem.\par

The combination of the \textit{Knapsack Random Set Theorem}, \textit{Knapsack Composition Theorem}, \textit{Knapsack M Quality Reduction Theorem}, and \textit{Knapsack Set Quality Theorem} indicate that any deterministic polynomial time solution to the \textit{Knapsack} problem, which relies upon some quality of the problem, will not produce a solution that works for all instances of the \textit{Knapsack} problem.\par

The \textit{K-SAT Input Relation Theorem} indicates that a solution dependant on the form of any \textit{NP-Complete} problem, or some quality of the problem's inputs, to sidestep the \textit{P = NP Optimization Theorem} will not produce a fast solution for the underlying K-SAT.  It is then the case that such an optimization is not transferable to all other \textit{NP-Complete} problems.\par
\vspace{4ex}

\noindent It has therefore been determined that:
\begin{enumerate}
	\item Search algorithms will not prove $P$ = \textit{NP}.
	\item Partitioning the search group will not produce an algorithm that proves $P$ = \textit{NP}.
	\item Relying upon the form of an individual \textit{NP-Complete} problem, or the quality of the problem's inputs, can sometimes solve that problem in deterministic polynomial time but does not provide an algorithm that proves $P$ = \textit{NP}.
	\item There is no magical solution that produces a deterministic polynomial time solution for \textit{NP-Complete} problems.
\end{enumerate}
\vspace{1ex}

\noindent If it can be accepted that any algorithm that solves a \textit{NP-Complete} problem will be one of the following:
\begin{enumerate}
	\item a search algorithm, which examines all possible literal values;
	\item a partitioned search algorithm, which partitions the set of all possible literal values and only examines a limited number of the possible input sets;
	\item a direct solution relying upon the form of the problem;
	\item or a magical solution.
\end{enumerate}

Then the conclusion $P \neq$ \textit{NP} can be accepted as final.

\begin{flushright}
Q.E.D.
\end{flushright}

\section{Version history.}
The author wishes to encourage further feedback which may improve, strengthen, or perhaps disprove the content of this article. For that reason the author does not publish the names of any specific people who may have suggested, commented, or criticized the article in such a way that resulted in a revision, unless premission has been granted to do so.\par

\noindent \textbf{arXiv Current Version}\newline
3Sep08 Submitted to arXiv.
\begin{itemize}
	\item Minor changes.
\end{itemize}

\noindent \textbf{arXiv Version 4}\newline
29Aug08 Submitted to arXiv.
\begin{itemize}
	\item Minor error corrections.
\end{itemize}

\noindent \textbf{arXiv Version 3}\newline
27Aug08 Submitted to arXiv.
\begin{itemize}
	\item Addition of formal proof that base conversion is \textit{NP-Complete}.
	\item Addition of argument reguarding magical solutions.
\end{itemize}

\noindent \textbf{arXiv Version 2}\newline
25Aug08 Submitted to arXiv.
\begin{itemize}
	\item Addition of formalized argument.
\end{itemize}

\noindent \textbf{arXiv Version 1}\newline
23Aug08 Submitted to arXiv.

\begin{received}
Received xx/2008; revised xx/2008; accepted xx/2008
\end{received}

\begin{thebibliography}{Horowitz and Sahni 1974}

	\bibitem[Horowitz and Sahni 1974]{horowitz} \textsc{Horowitz, E.} and \textsc{Sahni, S.} 1974. Computing partitions with applications to the knapsack problem. \textit{J. Assoc. Comput. Mach}. Zbl 0329.90046.

	\bibitem[Karp 1972]{karp} \textsc{Karp, R.} 1972. Reducibility among combinatorial problems. \textit{Complexity of computer computations}, \textit{Proc. Sympos., IBM Thomas J. Watson Res. Center}. MR378476 Zbl 0366.68041.

	\bibitem[Meek Article 1 2008]{meek1} \textsc{Meek, J.} 2008. \textit{P} is a proper subset of \textit{NP}. \textit{arXiv:0804.1079 Article 1 in series of 4}.

	\bibitem[Meek Article 2 2008]{meek2} \textsc{Meek, J.} 2008. Analysis of the deterministic polynomial time solvability of the \textit{0-1-Knapsack} problem. \textit{arXiv:0805.0517 Article 2 in series of 4}.

\end{thebibliography}
\end{document}